\documentclass[journal]{IEEEtran}
\ifCLASSINFOpdf
\else
\fi
\usepackage[pdftex]{graphicx}
\usepackage{caption}
\usepackage{subcaption}

\usepackage{balance}
\usepackage[table]{xcolor}
\definecolor{light-gray}{gray}{0.79}
\definecolor{lighter-gray}{gray}{0.95}

\usepackage{algorithm}
\usepackage{algorithmicx}
\usepackage{algcompatible}
\usepackage{algpseudocode}
\usepackage{esvect}
\usepackage{booktabs}
\usepackage{float}
\usepackage{makecell}
\usepackage{textcomp}
\newcommand{\framework}{NETRA}

\begin{document}
%
\title{\framework: Enhancing IoT Security using \\NFV-based Edge Traffic Analysis}
\author{\IEEEauthorblockN{
Rishi Sairam\IEEEauthorrefmark{1}, 
Suman Sankar Bhunia\IEEEauthorrefmark{2},
Vijayanand Thangavelu\IEEEauthorrefmark{1},
Mohan Gurusamy\IEEEauthorrefmark{1}
\\}
\IEEEauthorblockA{\IEEEauthorrefmark{1}Department of Electrical and Computer Engineering, National University of Singapore\\}
\IEEEauthorblockA{\IEEEauthorrefmark{2}School of Computing, National University of Singapore\\}
Email:  rishi.sairam@u.nus.edu, suman@nus.edu.sg, vijay@u.nus.edu, gmohan@nus.edu.sg
}
\maketitle

\begin{abstract}
This is the era of smart devices or \emph{things} which are fueling the growth of Internet of Things (IoT). It is impacting every sphere around us, making our life dependent on this technological feat. It is of high concern that these smart things are being targeted by cyber criminals taking advantage of heterogeneity, minuscule security features and vulnerabilities within these devices. Conventional centralized IT security measures have limitations in terms of scalability and cost. Therefore, these smart devices are required to be monitored closer to their location ideally at the edge of IoT networks.
In this paper, we explore how some security features can be implemented at the network edge to secure these smart devices. We explain the importance of Network Function Virtualization (NFV) in order to deploy security functions at the network edge. To achieve this goal, we introduce \framework -- a novel lightweight Docker-based architecture for virtualizing network functions to provide IoT security. Also, we highlight the advantages of the proposed architecture over the standardized NFV architecture in terms of \emph{storage, memory usage, latency, throughput, load average, scalability} and explain why the standardized architecture is not suitable for IoT. We study the performance of proposed NFV-based edge analysis for IoT security and show that attacks can be detected with more than 95\% accuracy in less than a second.

\end{abstract}

\begin{IEEEkeywords}
IoT, Edge, Security, NFV, Docker.
\end{IEEEkeywords}

%
\IEEEpeerreviewmaketitle

\section{Introduction}
\label{sec:intro}
Internet of Things (IoT) is a technology that facilitates interaction between physical and digital world. It is comprised of sensors and other electronic equipments that are remotely controlled using the existing network infrastructure. A recent Gartner report~\cite{gartner} says, by 2021, \$2.5 million per minute will be spent on these devices. IoT generates tonnes of data that can be extracted to give better quality of life by providing value-added services such as home automation, health care, industry automation etc. Though it is expected to change our lives rapidly, with huge volume of data on the network, it is exposed to general security threats such as denial of service (DoS) attacks and other cyber attacks. Thus, an important question is : how secure are these IoT devices? 

Recent attacks like Mirai \cite{mirai}, IoTroop and Stuxnet answer the above question very well. These attacks focused to make use of vulnerabilities in IoT devices itself. With the emergence of low cost devices and cut-throat competitions, manufacturers rush these IoT devices to the market almost without any security features or with very minimal ones. Also, these low cost tiny resource-constrained devices are not fit to use conventional IT security measures. Hence, there is a pressing need to improve security in the IoT environment. Lots of research has been done on how to make the current security systems more robust and viable to take care of IoT \cite{sood2016}. As the IoT device level security is erratic, incorporation of security features at network edge may help securing these devices. In this paper, we focus to push our research towards the network edge i.e. more closer to the IoT devices.  

Due to the massive number of heterogeneous devices, defining static rules for security purposes will not help. Instead, individual traffic analysis of these devices is required to understand their usual pattern and anomalous traffic. If this analysis is done in a centralized manner, scalability issues will arise. Thus, pushing the traffic analysis to the network edge will be helpful. Machine learning is used to cater various security needs of today's networks~\cite{xiao2018iot}. In this paper, we make use of machine learning algorithms to perform traffic analysis at the network edge, thus deeming it useful for the IoT environment.

At the network edge, usually we find devices with lower computational capabilities such as IoT gateways, home routers etc. The challenge is to implement security features for IoT at these edge infrastructure. For this purpose, Network Function Virtualization (NFV) can be used. NFV is a technique by which we can virtualize the traditional network functions using their software counterparts. The use of NFV technology to provide IoT security has been attracting the attention of researchers recently. The number of companies that are pushing the research and development of NFV paradigm is growing steadily since NFV can improve cost efficiency, flexibility, performance and reduce the capital expenditures (CAPEX) and operational expenditures (OPEX). 

The network functions that are implemented are called as \emph{virtual network functions} or VNFs and are usually deployed on high-capacity servers or cloud infrastructure. Given the low computational requirement at the network edge i.e. IoT gateway, we need a virtualization technique that is lightweight and scalable~\cite{natarajan2017analysis}. As an alternative, Docker containers promise one such environment, which open up opportunities for the security functions to be implemented as Docker VNFs.

\begin{table*}[t]
\caption{Comparison of Docker and Virtual Machines}
\begin{center}
\begin{tabular}{|l|l|l|}
\hline
\textbf{Feature} & \textbf{Docker} & \textbf{Virtual Machines} \\
\hline
Start time & $<$50 ms & 30 - 45 seconds \\
\hline
Stop time & $<$50 ms & 5 - 10 seconds \\
\hline
System Overhead & No overhead & Overhead due to hypervisor \\
\hline
Storage space & Tens of MBs in size & Tens of GBs in size \\
\hline
Scalability & Highly scalable & Not easily scalable \\
\hline
CPU load when idle & Normal & $\sim$1.5\% more than docker \\
\hline
Isolation & Less isolation due to software virtualization & More isolation due to hardware virtualization \\
\hline
Network round trip latency & $\sim$75 $\mu$s & $\sim$60 $\mu$s\\
\hline
I/O throughput (Read and Write) & 100000 I/Os per second & 50000 I/Os per second\\
\hline
\end{tabular}
\end{center}
\label{tab:comparison}
\vspace{-0.19in}
\end{table*} 

\begin{figure}[h]
\centering
\includegraphics[scale=0.48]{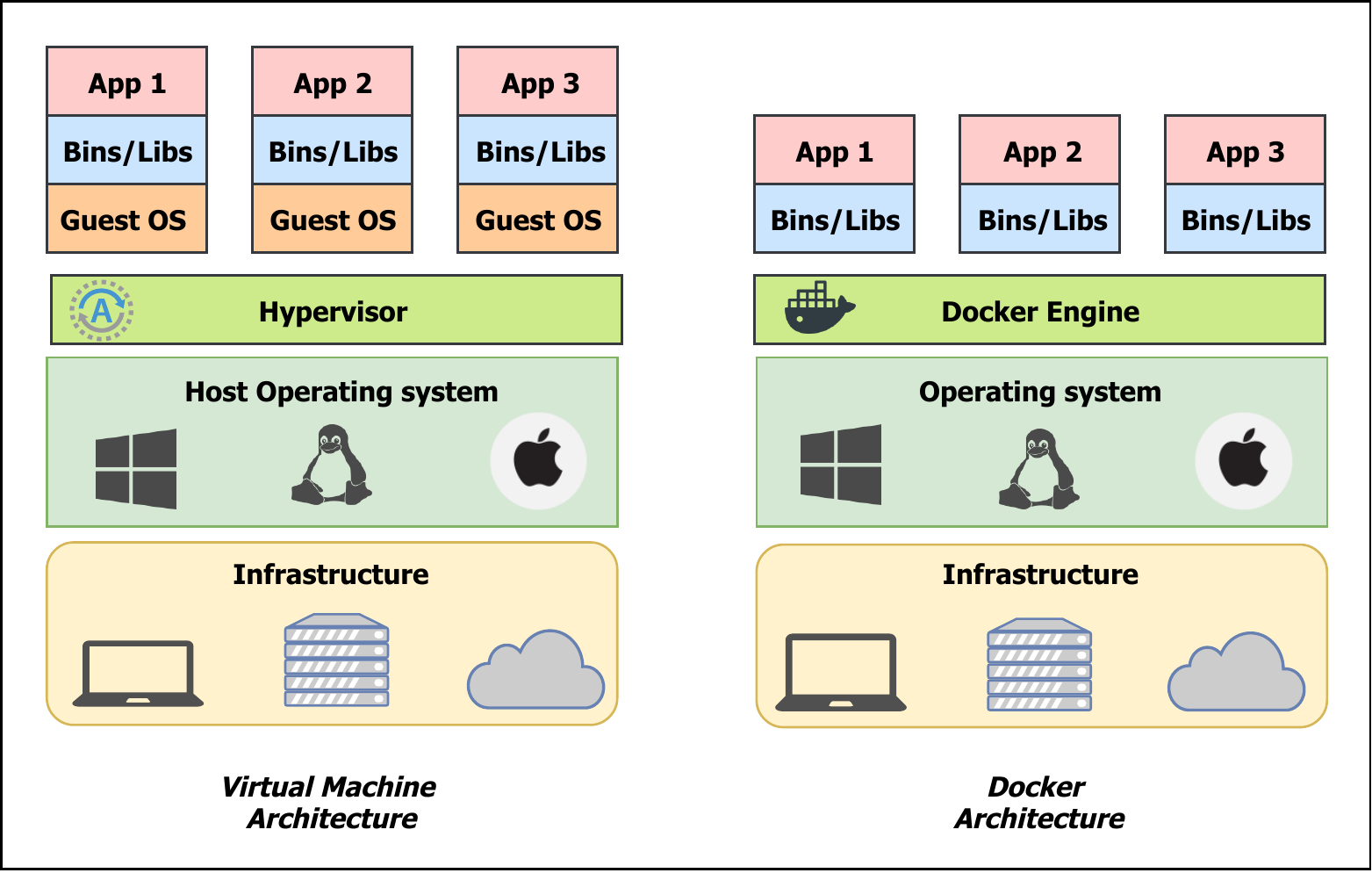}
\caption{Architectural Comparison of VMs and Dockers.}
\label{comparison}
\vspace{-0.1in}
\end{figure}

A Docker container is a stand-alone and light-weight executable package of software that includes everything needed to run it: code, settings, system libraries and system tools~\cite{docker}. Containers isolate the software running in it from its surroundings thus giving us a software level isolation unlike the hardware level isolation of virtual machines (VMs). Architecture of both VMs and Dockers are shown in Fig.~\ref{comparison}. 

In Fig.~\ref{comparison}, we can see that the hypervisor in VM, is replaced by the docker engine in containers. Containers are smaller than VMs that enable faster start up with better performance, less isolation and greater compatibility due to sharing of the host's kernel. Also, huge disk space can be saved as bulky guest operating systems for each application, are not required. Each application can be built into a docker image and hosted in cloud. It results in zero building time as the same docker image can be used to spawn many docker instances of the same application. The other key differences between VMs and Dockers are shown in Table \ref{tab:comparison}. 

In this paper, \framework -- a novel Docker based architecture is introduced for virtualizing network functions at IoT gateway. Here, each network function can be fetched from the cloud and deployed immediately onto the IoT gateway. Using the architecture, we implement VNFs that can improve the security of IoT environment and test using IoT devices like smart cameras, smart sockets etc. We collect real time traffic from a TP-Link smart camera to train our \emph{edge-analysis} VNF discussed in Section \ref{sec:edge_traffic}. Our models are able to successfully detect attacks with $\sim$95\% accuracy in nearly a second.

The rest of the paper outlines the way to harness the power of Dockers and NFV to provide a secure platform for IoT. First, differences between the existing VM-based platform and the proposed Docker-based architecture are discussed. Later, performance evaluation is presented. Finally, we discuss related works in this domain before making concluding remarks.

\section{Architecture}
\label{sec:architecture}

In the domain of network function virtualization, number of research and development activities are going on to provide a robust and efficient platform. One such standardized platform is the Open Platform for Network Function Virtualization or OPNFV~\cite{OPNFVwebsite}. It is patronized by leading companies and organizations. OPNFV provides a platform for the development and evolution of many NFV components that can be deployed in various open source ecosystems. In this paper, we propose NETRA -- a new architecture for deploying virtual network functions in low computational devices at the network edge like the IoT gateway. In order to compare, we have implemented both the architectures.
In this section, we describe the existing VM-based architecture (OPNFV) and our proposed architecture using Docker containers (NETRA).
\vspace{-0.05in}
\subsection{VM-based Architecture}

\begin{figure}[h]
\centering
\includegraphics[height=2.5in,width=0.9\linewidth]{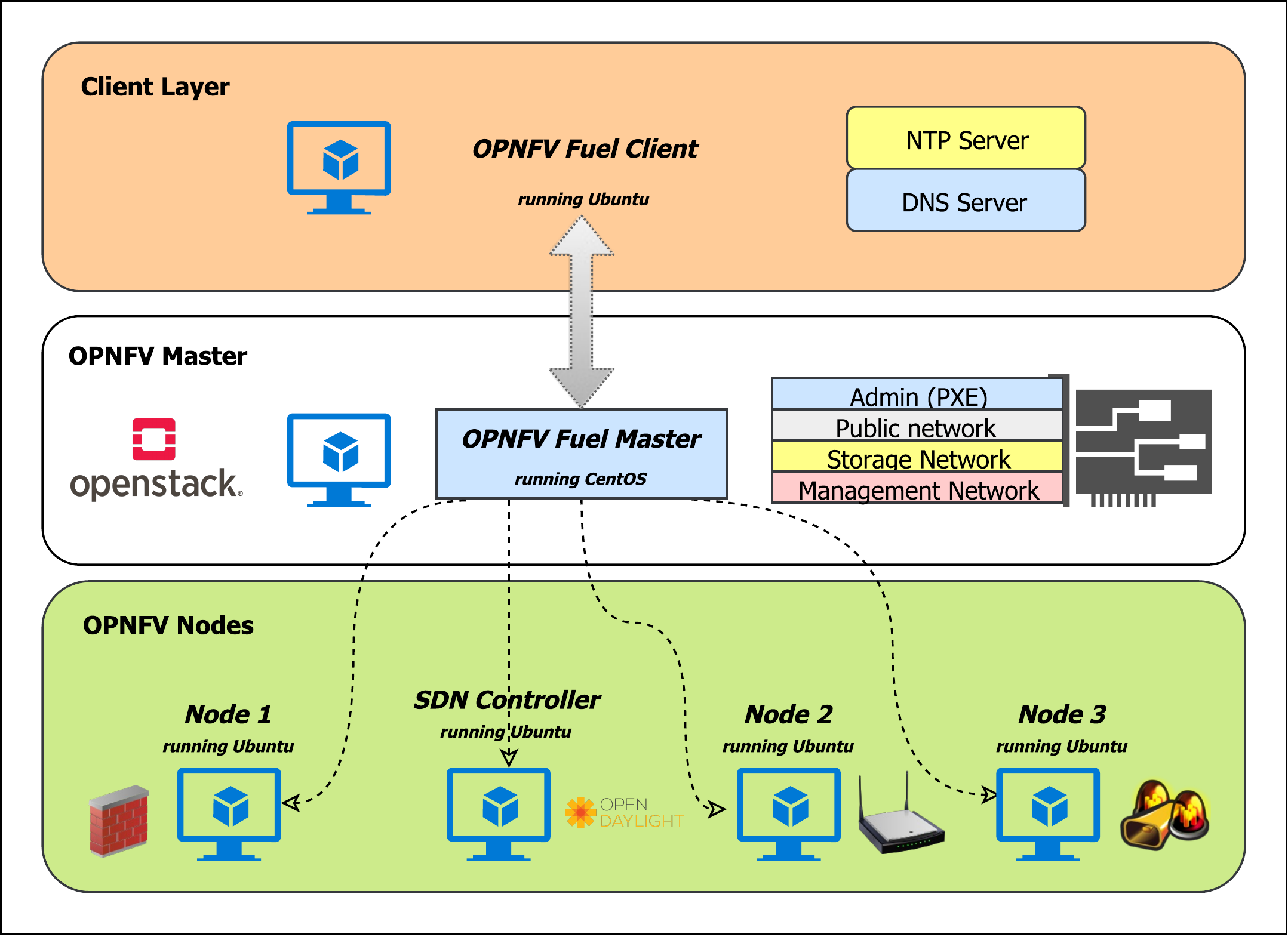}
\caption{OPNFV based OpenStack deployment of NFVs.}
\label{OPNFV_architecture}
\end{figure}

OPNFV (Open Platform for NFV) is basically an open-stack environment for deploying various VNFs using VMs. OPNFV provides various installers like \emph{Compass, Fuel} etc., for deploying the same. Because of the ease of use, Fuel~\cite{opnfvwiki} has been selected and 5 VMs are created to set up the environment. Fig.~\ref{OPNFV_architecture} shows the high-level view of the environment based on the OPNFV architecture.  We describe each of the three layers in the architecture below.

\subsubsection{OPNFV Master}
This is the layer where the OPNFV master is deployed. In our case, it is the Fuel Master, which is responsible for deploying the OPNFV nodes in the bottom layers. As shown in Fig.~\ref{OPNFV_architecture}, the master has a dedicated network card for each of the operations like administration, management, storage etc. It is important to note that this layer does not appear at the network edge rather it resides one layer above the nodes. 

\subsubsection{OPNFV Nodes}
This layer is the crux where the VNFs are deployed and it depicts the network edge in IoT scenario. We deployed 3 nodes or PODs (Point of Devices) that act as various elements of a NFV-based IoT environment. One VM is running \texttt{OpenDayLight} controller and the other two nodes are emulating different network functions (\emph{WAP} and \emph{firewall}). One of the biggest problems with an open stack deployment is the need for \texttt{PXE} booting ability, which is very unlikely to be present in the devices at this layer. Another noteworthy point is that, each of the nodes acts as a VNF, which implies that multiple VMs are required at the network edge.

\subsubsection{Client Layer}
The client layer is at the user end wherein the Fuel client can be accessed using a web interface. Using the Fuel client, one can deploy various VNFs with the help of the OPNFV master. 
In our experiment, we deployed the nodes detailed above with the help of the \texttt{OpenStack} interface through the Fuel client. 

In~\cite{cziva2017container}, it is discussed that IoT gateways have nearly the same specifications as that of a Raspberry Pi 3. Therefore, we model the network edge device with the help of a Raspberry Pi in this deployment. During the deployment, we noticed few shortcomings of OPNFV which renders it not suitable for use at network edge. The deployment of OPNFV demands at least 3 nodes (VMs) for deployment. This may be useful at an enterprise level environment. But it is futile to employ such an architecture at the edge of IoT network because the edge devices (i.e. IoT gateways) are of limited resource. 

As explained later in Section \ref{sec:evaluation}, this approach is not beneficial for the network edge scenario. Typically, the VMs in the OPNFV architecture require high computational abilities that cannot be expected at the network edge. Also, the deployment times are also very slow with OPNFV. Though OPNFV provides a standardized approach for deploying VNFs, it is not feasible enough to be used at the edge of IoT network.

\vspace{-0.1in}
\subsection{Proposed Docker-based architecture}

\begin{figure}[!h]
\centering
\includegraphics[height=2.5in,width=\linewidth]{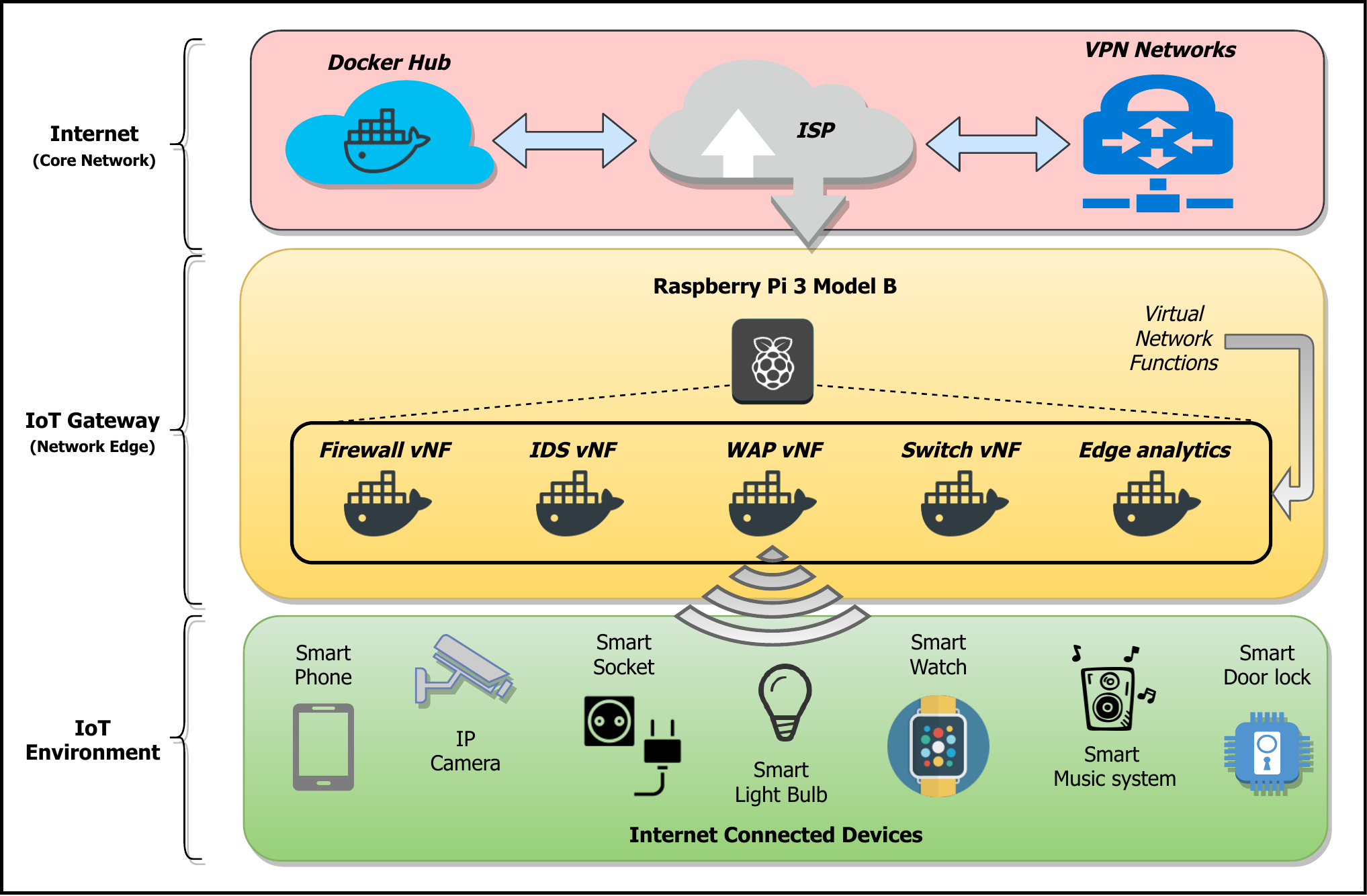}
\caption{Proposed Docker-based architecture with a Raspberry Pi at network edge.}
\label{proposed_docker}
\vspace{-0.1in}
\end{figure}

Having said about the issues faced by OPNFV, we require a robust and efficient architecture that can facilitate the deployment of VNFs at the edge of the network. 
In this section, we present our Docker-based architecture -- NETRA for NFV-based Edge TRaffic Analysis. The proposed architecture helps in creating and deploying VNFs in low computational devices. Fig.~\ref{proposed_docker} shows a high-level view of our proposed architecture. Layers of the architecture are discussed in detail below.  

\subsubsection{Core Network}
In this layer, connections to the cloud servers are routed through the Internet Service Provider (ISP). When IoT devices in the lowest layer, are activated, they talk with their respective cloud servers 
viz. \emph{D-Link} devices talk with \emph{D-Link cloud}. In this layer, we have the Docker Hub, which is a repository for all the Docker images that we have created. The Docker images can be deployed from this layer with a \texttt{docker pull} command. When a new VNF has to be deployed, it only takes a couple of minutes to build and run the Docker VNFs.

\subsubsection{IoT Gateway}
This layer represents the actual network edge of the IoT scenario. In our implementation, we use a Raspberry Pi 3 as the IoT gateway, which hosts various Docker VNFs. The IoT devices connect to the network that is created using the Wireless Access Point (WAP) VNF described in Section \ref{sec:vnf}. The communication between the IoT devices and external world go through this layer. Thus attacks can be detected and mitigated at this layer. As seen in the Fig.~\ref{proposed_docker}, we deploy a chain of multiple VNFs with the Raspberry Pi designated as an IoT gateway. This chain of VNFs work in tandem. We deploy security VNFs like firewall, intrusion detection, software defined networking (SDN) switch, edge analytics etc. as described in Section \ref{sec:vnf}.

\subsubsection{IoT Environment}
In this layer, we have the smart \emph{things} or the IoT devices that are connected to the Internet through the above layers. In our lab, we use devices like smart bulbs, IP cameras, smart remote controllers, smart sockets etc. 

The architecture can be further extended with an SDN controller managing the Open vSwitch running on the IoT gateway as a VNF. As discussed in earlier sections, containers are light-weight and can be used to run any application with a very low overhead and faster deployment times. NETRA also addresses the core issue of OPNFV which requires at least three nodes at the edge. Here all the VNFs are on the same device, yet isolated. The performance comparisons between the two architectures in Section~\ref{sec:evaluation} validate the fact that dockerizing VNFs is an effective way for bringing NFV to edge of the network.
\vspace{-0.1in}
\subsection{Virtual Network Functions for IoT Security}
\label{sec:vnf}
With the comprehensive analysis of the two architectures, it is evident that for an IoT environment, Docker-based VNFs are the best suitors. We implement some VNFs, those when coupled together, form an IoT gateway that is secure by itself and is lightweight at the same time. The details about the implemented VNFs are explained below.
\subsubsection{WAP as VNF}
In order to serve as an IoT gateway, the Raspberry Pi must be able to connect to the devices and forward the packets to Internet. To facilitate this, the WAP VNF creates a wireless access point for the IoT devices to connect using the \emph{hostapd} package and also acts as a DHCP server by running \emph{udhcpd}. Since the Raspberry Pi is connected to the Internet through Ethernet port, a simple masquerading rule will help in forwarding the packets.
\subsubsection{Firewall}
One of the most important yet simplest security function that any system demands is the firewall. Using this VNF, we create firewall rules to block a malicious user. The firewall rules are applied with the help of the  \emph{iptables} package. In case the user needs to be unblocked, we just need to stop the docker instance corresponding to that user's MAC.
\subsubsection{IDS}
Intrusion Detection Systems (IDS) are generally heavy on the processor and memory due to their large rule sets and continuous traffic analysis. In order to create a light weight version of IDS, we use minimalistic rules pertaining to the IoT environment on top of \emph{snort} package running with low memory option. When the IDS VNF detects a malicious activity, it instructs the WAP VNF to kick out the malicious user and spawns the firewall docker VNF to block the user.
\subsubsection{Software Defined Switch}
\label{ovs}
The Software Defined Switch is an advanced VNF added to the IoT gateway. This VNF uses \emph{Open vSwitch} and harnesses the power of software defined switches in managing the flows in and out of the network.
\subsubsection{Edge Analytics}
This VNF performs traffic analysis at the network edge using machine learning algorithm as discussed in Section \ref{sec:edge_traffic} for detecting and mitigating malicious attacks. The detection phase comprises of two stages : \emph{anomaly detection} and \emph{attack detection}. When the incoming traffic is detected as anomalous, the traffic is analyzed to ascertain the type of attack. After successful detection, we mitigate the attack using the software defined switch mentioned above. 
\vspace{-0.05in}

\section{Edge Traffic Analysis}
\label{sec:edge_traffic}
In this section, we discuss about the edge traffic analysis using NFV and machine learning (ML). We encapsulate the algorithms discussed below in an \emph{Edge Analytics} VNF 
at the IoT gateway itself.
We first extract the features for training our ML models so that we can detect anomalies in real time. The \emph{scapy} library in Python is used for performing the feature extraction using Algorithm~\ref{alg:feature}. 
We propose a majority vote type detection wherein three models detect whether the given datapoint comes from a malicious packet or not using Algorithm \ref{alg:anomaly}. Once an anomaly is detected, the attack detection module is spawned to identify whether it is a known attack and proper mitigation is achieved through the software defined Open Virtual Switch (OVS) VNF. 

\vspace{-0.1in}
\subsection{Feature extraction}

In order to have efficient detection, it is of utmost importance to have proper features to carry out the detection. Carefully selected features aid in creating an accurate predictive model. In our algorithm, the \emph{scapy} librabry is used for extracting the packets and features. For the training data, we connected a TPLink camera to the raspberry pi and collected packets in different scenarios: 1) with benign traffic flows, 2) while attacking the camera with SYN flooding and 3) while attacking the camera with ICMP flooding. SYN and ICMP flooding are carried out using \emph{hping3} tool found in the Kali Linux distro connected in the same network. \emph{Wireshark} tool is used to capture data in all the three scenarios and this datafile (\emph{.pcap file}) is given as the input to Algorithm~\ref{alg:feature}.

\begin{algorithm}[h]
\caption{Feature Extraction} 
\label{alg:feature}
 \hspace*{\algorithmicindent} \textbf{Input: } \textit{pcap, dev-IP, attack-IP} \\
 \hspace*{\algorithmicindent} \textbf{Output: } \textit{featurelist (or) datapoint}
\begin{algorithmic}[1]
    \If {online}
    	\State $pcap \Leftarrow $ sniff packet
        \State extract packet (\textit{pcap, dev-IP})
    \Else
    	\State extract packet (\textit{pcap, dev-IP})
        \State $sess \leftarrow pktlen + samp$
        \For {session}
        	\State $featurelist \Leftarrow extract features(pkt)$
            \If {IP == attack-IP}
            	\State $label \leftarrow X$ \Comment X represents the attack index
            \EndIf
        \EndFor
    \EndIf
\end{algorithmic}
\end{algorithm}

In Algorithm~\ref{alg:feature}, a sampling time is set to process the packets. This sampling time (\texttt{samp}) is expected to be in the order of seconds depending on the amount of packets at hand. When we have the pcap file, the training data gets generated and is called \emph{offline} feature extraction. Once we have a model, the algorithm switches to an \emph{online} mode, wherein real-time traffic is collected and extracted. The features are extracted in sessions that is dependent on the packet length (\texttt{pktlen}) and the previously set sampling time. Once the packet is extracted, we extract the features like \texttt{SYN} flag, \texttt{ACK} flag, \texttt{PUSH} flag, \texttt{URG} flag, protocol, mean of packet duration, variance of packet length, mean of arrival times, variance of arrival times, total number of packets etc. A total of 21 features are chosen out of many as the prediction was not affected by adding more features. The training data is labeled, whether it is clean or SYN flood or ICMP flood, with the help of the attacker IP and attacker protocol. During online feature extraction, the packets are filtered using the device's IP and protocol.
\vspace{-0.1in}

\subsection{Anomaly Detection}

Anomaly detection is the technique that helps us in identifying any unusual pattern that does not match with expected behavior. Outliers detection or anomaly detection can thus be used to find any malicious activity going on in the network, if we can model the usual activities~\cite{choudhary}.

\begin{algorithm}[h]
\caption{Anomaly Detection} 
\label{alg:anomaly}
 \hspace*{\algorithmicindent} \textbf{Input: } \textit{clean data, attacked data (or) datapoint} \\
 \hspace*{\algorithmicindent} \textbf{Output: } \textit{prediction}
\begin{algorithmic}[1]
    \If {training}
    	\State $X \leftarrow$ datapoints(\textit{clean csv})
        \State $X\_test \leftarrow$ datapoints(\textit{attacked csv})
        \State $\mathcal{M}1 \leftarrow IsolationForest(X)$
        \State $\mathcal{M}2 \leftarrow OneClassSVM(X)$
        \State $\mathcal{M}3 \leftarrow EllipticalEnvelope(X)$
        \State $\mathcal{P}1 \leftarrow \mathcal{M}1(X\_test)$
        \State $\mathcal{P}2 \leftarrow \mathcal{M}2(X\_test)$
        \State $\mathcal{P}3 \leftarrow \mathcal{M}3(X\_test)$
        \State $ \mathcal{P} \Leftarrow \mathcal{P}1 + \mathcal{P}2 + \mathcal{P}3$ \Comment Testing the model
    \Else
    	\State $\mathcal{P}1 \leftarrow \mathcal{M}1(datapoint)$
        \State $\mathcal{P}2 \leftarrow \mathcal{M}2(datapoint)$
        \State $\mathcal{P}3 \leftarrow \mathcal{M}3(datapoint)$
        \State $ \mathcal{P} \Leftarrow \mathcal{P}1 + \mathcal{P}2 + \mathcal{P}3$
    \EndIf
    \If {$\mathcal{P} < 0$}
    	\State \textbf{Anomaly Detected}
        \State attack\_detection(\textit{datapoint})
    \EndIf
\end{algorithmic}
\end{algorithm}

The training data is obtained by capturing packets when there is no malicious activity and is further used to train the models of the usual behavior. Three models are considered for detection, namely: 
\begin{itemize}
\item One-Class SVM : One of the popular algorithms that learns a decision function for outliers detection is One-Class Support Vector Machine (SVM) and is used to classify new data as either similar or anomalous. The decision rule of the One-Class SVM is given by Equation~\ref{SVM}~\cite{vlasveld}.
\vspace{-0.1in}
\begin{equation}\label{SVM}
    f(x) = sgn(\sum_{i=1}^n \alpha_i y_i K(x,x_i) + b)
\end{equation}

where $\texttt{sgn}(\cdot)$ is the sign function, $\alpha_i$ are the Lagrange functions that support the machine and $\texttt{K}(\cdot)$ is the kernel function. In our model, the \emph{radial base function} (RBF) is chosen as the kernel.\\
\vspace{-0.1in}
\item Isolation Forest : Instead of constructing a model for the normal activities (packets in our case) and then identifying the outliers, Isolation Forest follows a model-based method that explicitly isolates anomalies. It does that through an algorithm that has a linear time complexity with a low constant and a low memory requirement~\cite{liu_ting_zhou_2008}. The algorithm calculates an anomaly score (given by Equation~\ref{IF}) to distinguish between normal cases and anomalies.
\begin{equation}\label{IF}
    s(x, n) = 2^{-\frac{E[h(x)]}{c(n)}}
\end{equation}
where $E[h(x)]$ is the average path length of a point x in the tree and $c(n)$ is the average path length of an unsuccessful search in the binary search tree.\\
\vspace{-0.1in}
\item Elliptical Envelope : The Elliptical Envelope package uses the covariance between the features to detect anomalies in a Gaussian distributed dataset. It basically tries to find an elliptical boundary that can hold most of the data. Any data that falls outside this boundary is classified as anomalous. This model uses the FAST-Minimum Covariance Determinate method \cite{doi:10.1080/00401706.1999.10485670} to find the size and shape of the ellipse with the help of the Mahalanobis distance shown in Equation \ref{MD}.
\begin{equation}\label{MD}
    d_{MH} = \sqrt[]{(\vec{x} - \vec{\mu})^TC^{-1}(\vec{x} - \vec{\mu})}
\end{equation}
where $\vec{x}, \vec{\mu}$ is the data vector and its mean. Equation \ref{MD} basically measures the deviation of a data vector from its mean.\\
\end{itemize}
\vspace{-0.1in}
For testing the model, we generate a simple SYN flood attack on the IoT device and extract features using Algorithm~\ref{alg:feature}. Our majority voting model has helped in reducing the number of false positives and missed detections. If the model detects an anomaly, it returns -1 and if not, returns 1. As shown in Algorithm~\ref{alg:anomaly}, we predict whether the test data is malicious or benign using all the three models and decide using majority votes. This way we harness the advantages of all the three models combined to get efficient detection. This algorithm runs inside a docker VNF in the IoT gateway to detect anomalies (malicious activities) in the environment. In order to detect the attack and block the malicious user, we call the \emph{attack detection} algorithm once an anomaly is detected.
\vspace{-0.05in}
\subsection{Attack Detection and Mitigation}

The attack detection module gets its input data point from the anomaly detection module, when the latter classifies it as being malicious. Once it gets its data point, it follows Algorithm~\ref{alg:attack} to detect the kind of attack and spawns the mitigation process if it is a known attack.

\begin{algorithm}[h]
\caption{Attack Detection}
\label{alg:attack}
 \hspace*{\algorithmicindent} \textbf{Input: } \textit{labeled\_data, datapoint} \\
 \hspace*{\algorithmicindent} \textbf{Output: } \textit{attack\_type}
\begin{algorithmic}[1]
    \If {training}
    	\State $t, \mathcal{T} \Leftarrow test\_train(labeled\_data)$
        \State $X\_train \leftarrow datapoints(\mathcal{T})$
        \State $Y\_train \leftarrow labels(\mathcal{T})$
        \State $X\_test \leftarrow datapoints(t)$
        \State $Y\_test \leftarrow labels(t)$
        \State $\mathcal{M} \leftarrow RandomForest(X\_train, Y\_train)$
        \State $ \mathcal{P} \Leftarrow \mathcal{M}(X\_test)$ \Comment Testing the model
        \State $\mathcal{E} = diff(\mathcal{P}, Y\_train)$ \Comment Classification Error
    \Else
    	\State $attack\_type \Leftarrow \mathcal{M}(datapoint)$
        \If {$attack\_type \neq $ NORMAL} \Comment If not benign
        	\State \textit{call OVS vNF}
        \EndIf
    \EndIf
\end{algorithmic}
\end{algorithm}

For this module, we took \emph{training data} while attacking the IoT device using \texttt{SYN} flooding and \texttt{ICMP} flooding and labeled the data accordingly using Algorithm~\ref{alg:feature}. Similarly, more attack signatures can be given. The model is tested with the same labeled data set by having a 60:40 split for training and testing respectively. We chose the \emph{RandomForest} classifier for its high accuracy and effective classification.  

Random Forest basically builds multiple decision trees and couples them all to form a stable and accurate prediction. Important reason for choosing Random forest is its ability to rank features based on importance, which we will observe in Section~\ref{perf}. Another advantage of this classifier is its randomness, due to which over-fitting problem is avoided~\cite{donges_2018}.

Algorithm~\ref{alg:attack} is thus quite straightforward in its operation. It detects whether the data point from the anomaly detection module is a known attack or not. If yes, it calls the Open vSwitch VNF and applies appropriate flow rules to block the malicious traffic. 
\vspace{-0.05in}


\begin{figure*}[!h]
\begin{subfigure}{.5\linewidth}
\centering
\includegraphics[width=\linewidth]{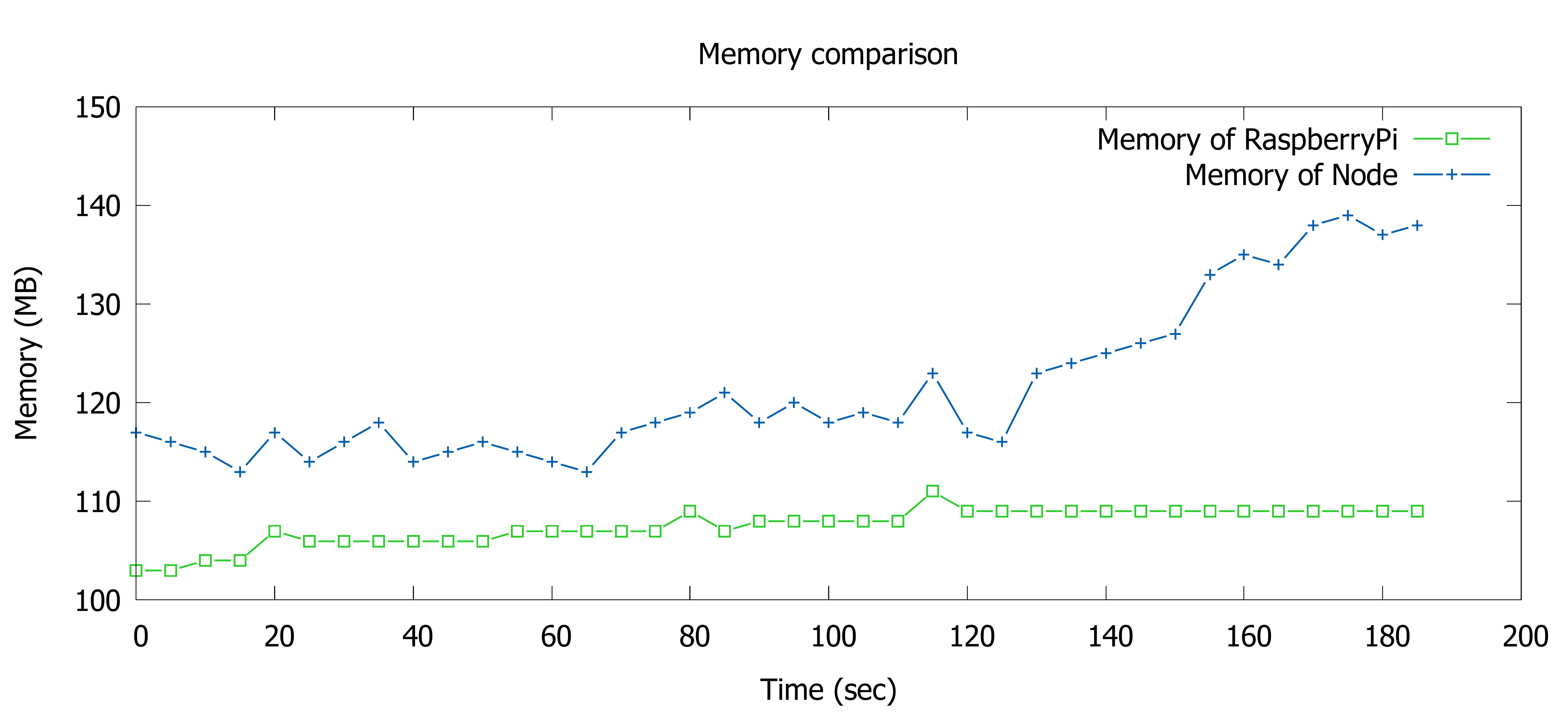}
\caption{Memory comparison between the two architectures.}
\label{fig:sub1}
\end{subfigure}%
\begin{subfigure}{.5\linewidth}
\centering
\includegraphics[width=\linewidth]{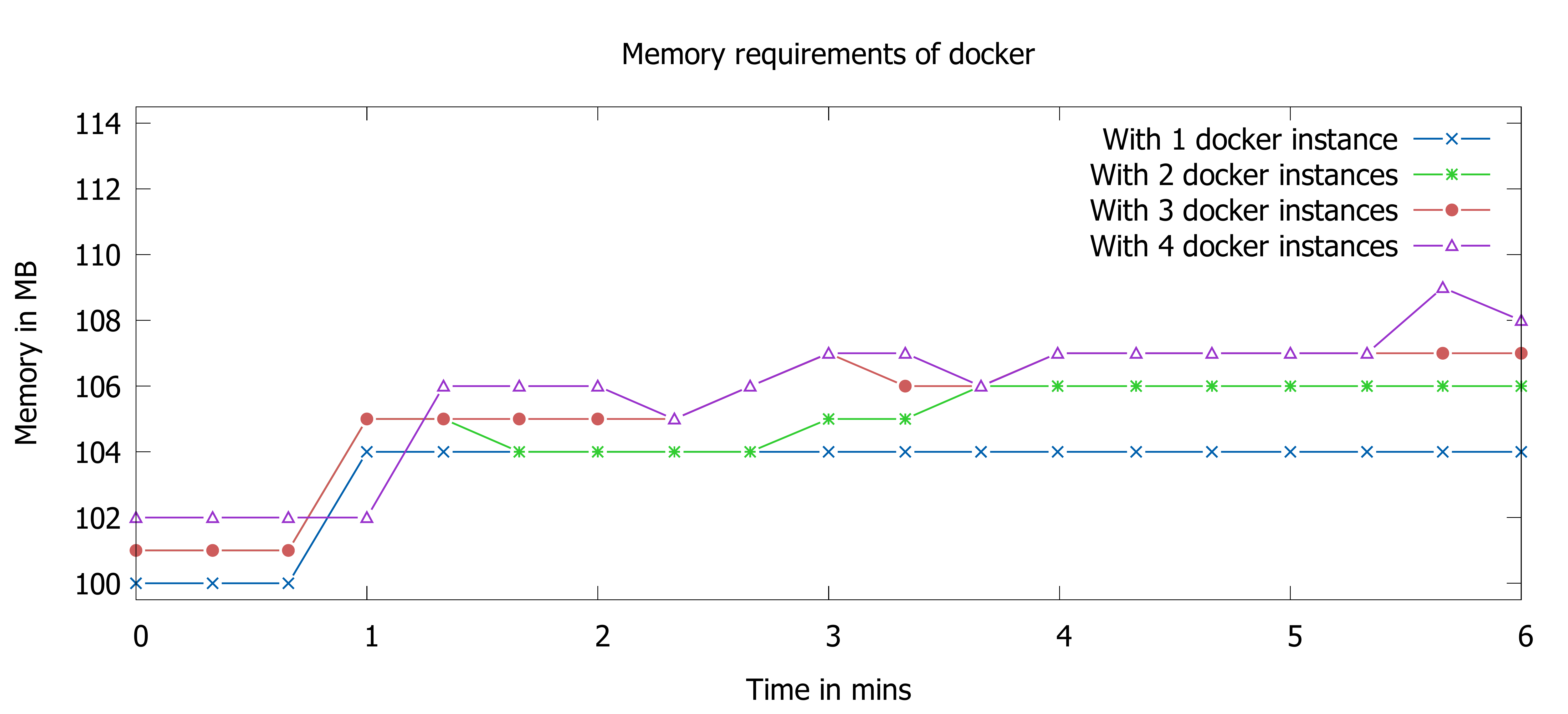}
\caption{Memory usage of multiple docker instances.}
\label{fig:sub2}
\end{subfigure}\\[1ex]
\begin{subfigure}{0.5\linewidth}
\centering
\includegraphics[width=\linewidth]{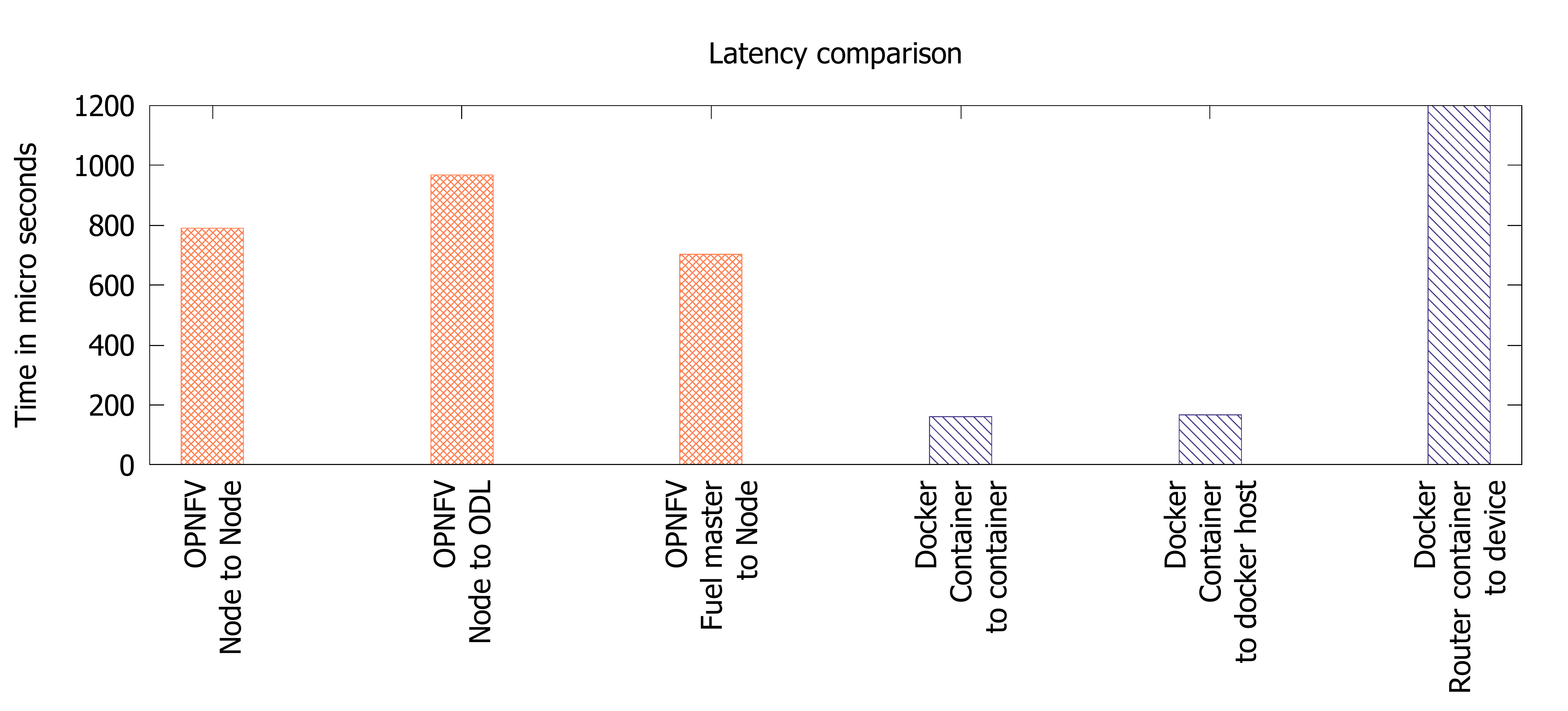}
\caption{Latency comparison based on ICMP.}
\label{fig:sub3}
\end{subfigure}
\begin{subfigure}{0.5\linewidth}
\centering
\includegraphics[width=\linewidth]{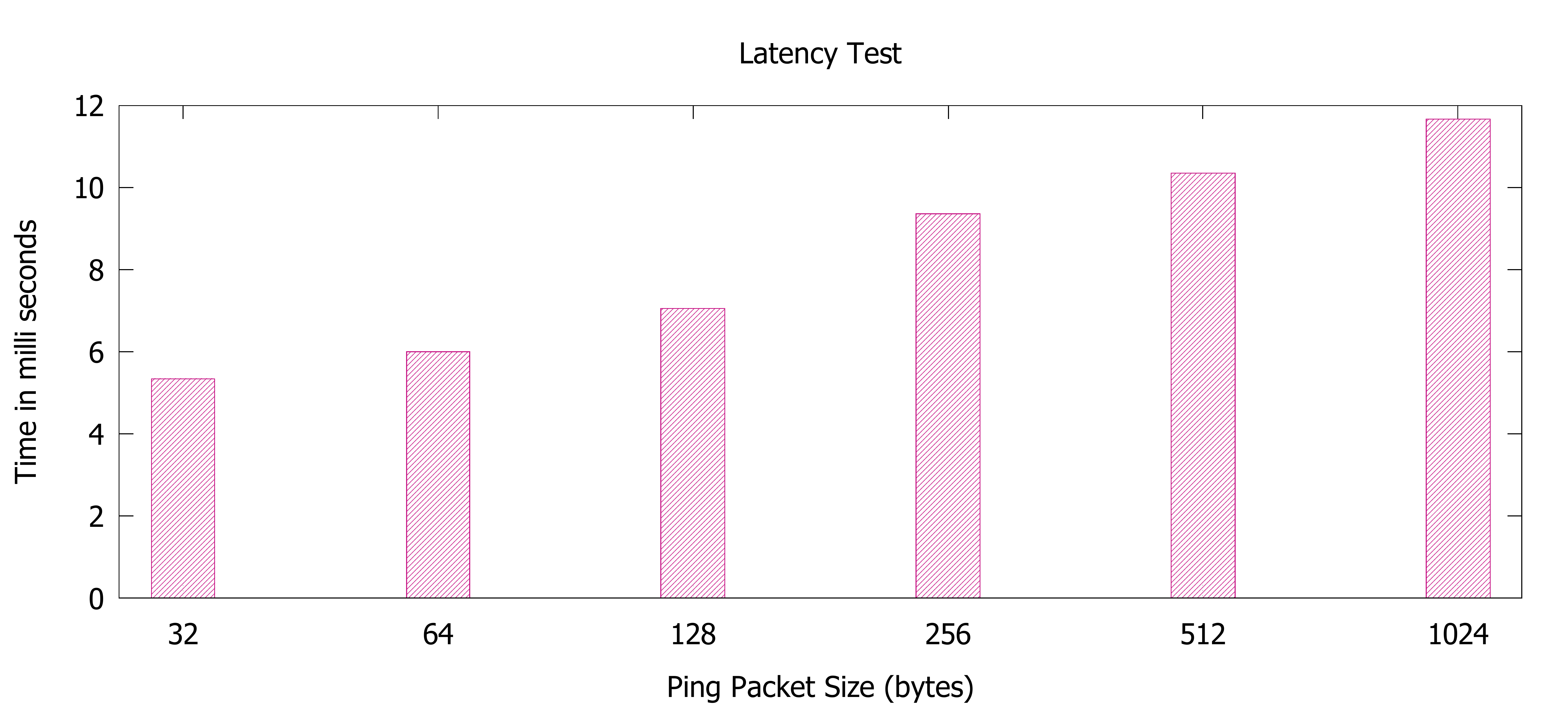}
\caption{Latency test between external host and docker containers using various sized ping requests.}
\label{fig:sub4}
\end{subfigure}
\caption{Memory requirements and Latency comparison.}
\label{fig:test}
\vspace{-0.1in}
\end{figure*}

\section{Performance Study}
\label{sec:evaluation}

\subsection{Efficiency of Proposed Architecture (\framework)}

In this section, we compare the two architectures - VM-based (OPNFV) and Docker-based (\framework) - in terms of key performance indicators such as storage, memory, latency, network and scalability~\cite{rfc8172}. Also, we discuss how a container based architecture helps us in deploying multiple VNFs in a single host/device. 

All these results are based on the experiments carried out in lab setup. For the OPNFV architecture, the VMs were built on top of a performance laptop having Intel i5 7th generation processor, 16 GB RAM and 2TB HDD. For \framework, the Raspberry Pi 3 Model B has a 1.2 GHZ quad-core ARMv8 Cortex A53 processor and a Broadcom IV GPU with 1 GB DDR2 RAM and in-built WLAN NIC.

\subsubsection{Storage}
In the case of OPNFV, there are minimum size requirements that need to be met before deploying an open stack environment. For a node that acts as the SDN controller (based on OpenDayLight) with Ceph storage, OPNFV demands 72GB storage. Similarly, a typical VNF node with just the operating system requires a minimum of 25GB storage, which is very unlikely for a network edge device.

In the case of \framework, we can have multiple VNFs on a Raspberry Pi which has only 16G of storage. Based on our experiments, a single Raspbian image takes up only 126MB, our WAP VNF image takes 182MB and our firewall VNF takes up 157M. The docker instances use still lower sizes with firewall instance taking 222B of storage and WAP instance taking only 18 KB of storage.

\subsubsection{Memory}
As mentioned in Section \ref{sec:architecture}, OPNFV requires the nodes to be PXE bootable. Based on our tests, PXE booting with \emph{Fuel}~\cite{opnfvwiki} required at least 1GB of RAM for each of the nodes. The controller node had higher memory requirement of 2GB for running the java based OpenDayLight controller. In our case, the Raspberry Pi 3 comes with only 1 GB of RAM. But due to the very low memory foot prints of docker, we are able to instantiate many VNF instances using the same.
    
The memory comparison of the two architectures is shown in Fig.~\ref{fig:sub1}. We observe that the memory of the node is high due to the OS, which shoots up even more after starting the \texttt{hostapd} service in the node. The peaks in the memory line indicate the start of a docker instance. It can be noted that each docker instance takes up only 1MB of RAM, which makes docker the best choice for VNFs at the IoT gateway.

Fig.~\ref{fig:sub2} shows the memory graphs of the Raspberry Pi as a host when multiple docker containers are launched. We see that the memory requirements of the docker instances are very low and differ in the order of only 1 MB with 4 different tests. The experiments were done by instantiating up to 4 docker instances simultaneously in repeated iterations. The peaks in the graphs explain the start of the docker instance, which later come down.

\subsubsection{Latency}
When working with a chain of VNFs, it is important to ensure that the latency between two VNFs is as low as possible. In this section, we compare the latency between different VNFs in both the architectures with the help of pinging the nodes through ICMP. Fig.~\ref{fig:sub3} shows that the latency between the nodes in OPNFV is higher as compared to the dockerized VNFs. This is due to the fact that, the docker VNFs are in the same host while the VNFs in OPNFV are in different VMs. We also note that the latency between the IoT device and WAP VNF is quite high (10.2 ms), this is because, the device is connected wirelessly to the WAP, where the losses are quite high.

We carry out experiments to measure the latency with different sizes of \texttt{ping} command. In Fig.~\ref{fig:sub4}, we observe that the latency is small on the order of milliseconds.


\subsubsection{Network}
Though the latency in NETRA is very less compared to OPNFV, the number of network interfaces is limited in our case, as the Raspberry Pi comes with only 2 NICs. In the case of OPNFV, the nodes have as many NICs as the VM can support.

With many NICs, we can have many services like administration, orchestration, management communicating without interference. But this comes with a trade-off, with more connections, each connection requires a higher bandwidth link, which is expensive. In \framework, high bandwidth between the VNFs is ensured since all the VNFs are practically on the same device. 

\begin{table}[!h]
\caption{Throughput tests with \emph{iperf} tool}
\label{tab:sub5}
\begin{tabular}{|l|l|l|}
\hline
\textbf{Configuration} & \textbf{Bandwidth \newline (Mb/s)} & \textbf{Buffer Size \newline (Kb)} \\
\hline
TCP (1 thread) & 16.8 & 43.8 \\
\hline
TCP (3 threads) & 5.6 & 43.8 \\
\hline
UDP & 1.05 & 160 \\
\hline
UDP with 100 Kbps bandwidth & 0.1 & 160 \\
\hline
\end{tabular}
\end{table}


To calculate the throughput performance of docker containers, we make use of the \emph{iperf} tool available in Linux platforms. As shown in Table~\ref{tab:sub5}, we carry out simple \emph{iperf} tests which involve data transfer between the \emph{iperf} client, the Docker instance and an \emph{iperf} server. A total of four tests are carried out, the first one involving a TCP connection, the second one with 3 parallel connections (the resulting throughput is per connection), the third one with an UDP connection which results in 2650 datagrams to be exchanged while the other test involving a constrained bandwidth of 100 kbps resulted in only 267 datagrams.

\subsubsection{Average Load}
\begin{figure}[!h]
\centering
\includegraphics[height=2in,width=0.98\linewidth]{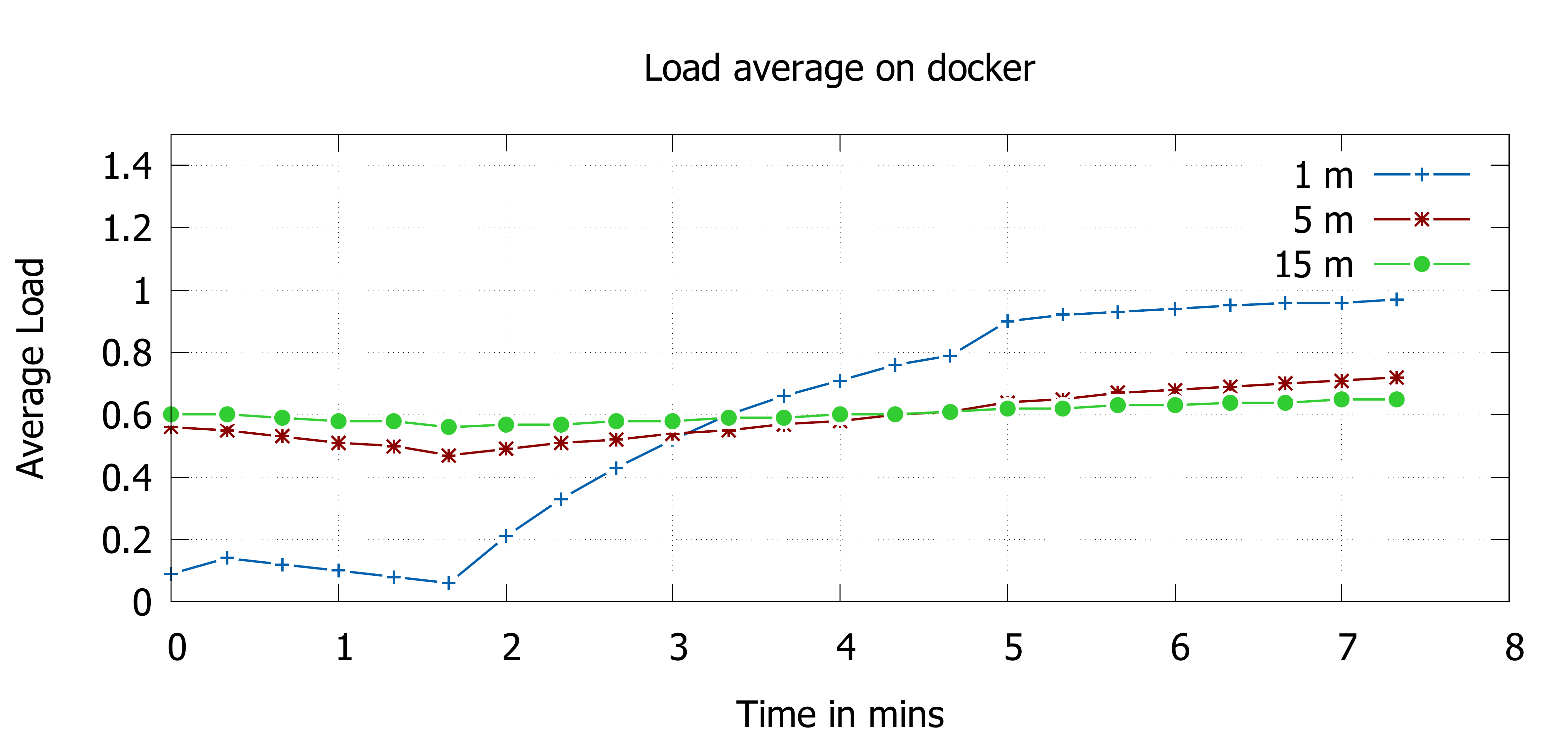}
\caption{Average load of the host from \emph{top} command.}
\label{fig:sub7}
\end{figure}

	Another important performance metric to be considered is the average load of the host running the dockers. The dockers should not use up the resources of the host during their working. With low performance hosts as in our case, it is very important to have a stable average load in the long run. Fig.~\ref{fig:sub7} shows the average load of the Raspberry Pi running 4 docker instances. We see that the 1-minute average increases with each docker instance, but the long-term averages remain stable, implying that the system is still stable. With these results, we can vouch that the Docker containers indeed consume very little resources and thus scale with even low performance edge devices.

\subsubsection{Scalability}

The last, but important performance metric is scalability. We can clearly see from the storage and memory requirements that NETRA is very much scalable. For example, we can have hundreds of firewall VNFs with individual rule sets running on a simple raspberry pi having very less computational strength.

Another important aspect of scalability is the deployment times. In order to be scalable, the Docker environment should also provide faster deployment time. In Table~\ref{tab:sub6}, we can see that the deployment time in running a Docker instance is as low as a second. Time for building the image for the first time, where it downloads everything from the cloud, is nearly 5 mins for our WAP VNF. In the case of OPNFV, we note that we cannot expect such short deployment time. This can be clearly seen from Table~\ref{tab:sub6} as bringing up a complete node takes nearly 15-20 minutes.
\begin{table}[!h]
\centering
\caption{Deployment Time for VNFs}
\label{tab:sub6}
\begin{tabular}{| p{5cm} | p{1.8cm} |}
\hline
\textbf{Process} & \textbf{Time taken} \\
\hline
OPNFV Bring up a VNF & $\sim 20$ minutes  \\
\hline
OPNFV Start/Stop a VNF & $\sim 5$ minutes  \\
\hline
HostAP Docker Time to build container for the first time & 2 minutes and 20 seconds \\
\hline
Firewall Docker Time to build container for the first time & 4 minutes and 16 seconds \\
\hline
HostAP Docker Time to build container & 9 seconds \\
\hline
Firewall Docker Time to build container & 5 seconds \\
\hline
HostAP Docker Time to run container & $<1$ second \\
\hline
Firewall Docker Time to run container & $<1$ second \\
\hline
\end{tabular}
\end{table}

\vspace{-0.2in}    

\subsection{Performance of NFV-based Edge Analysis for IoT Security}
\label{perf}
\begin{figure}[h]
\centering
\includegraphics[width=\linewidth]{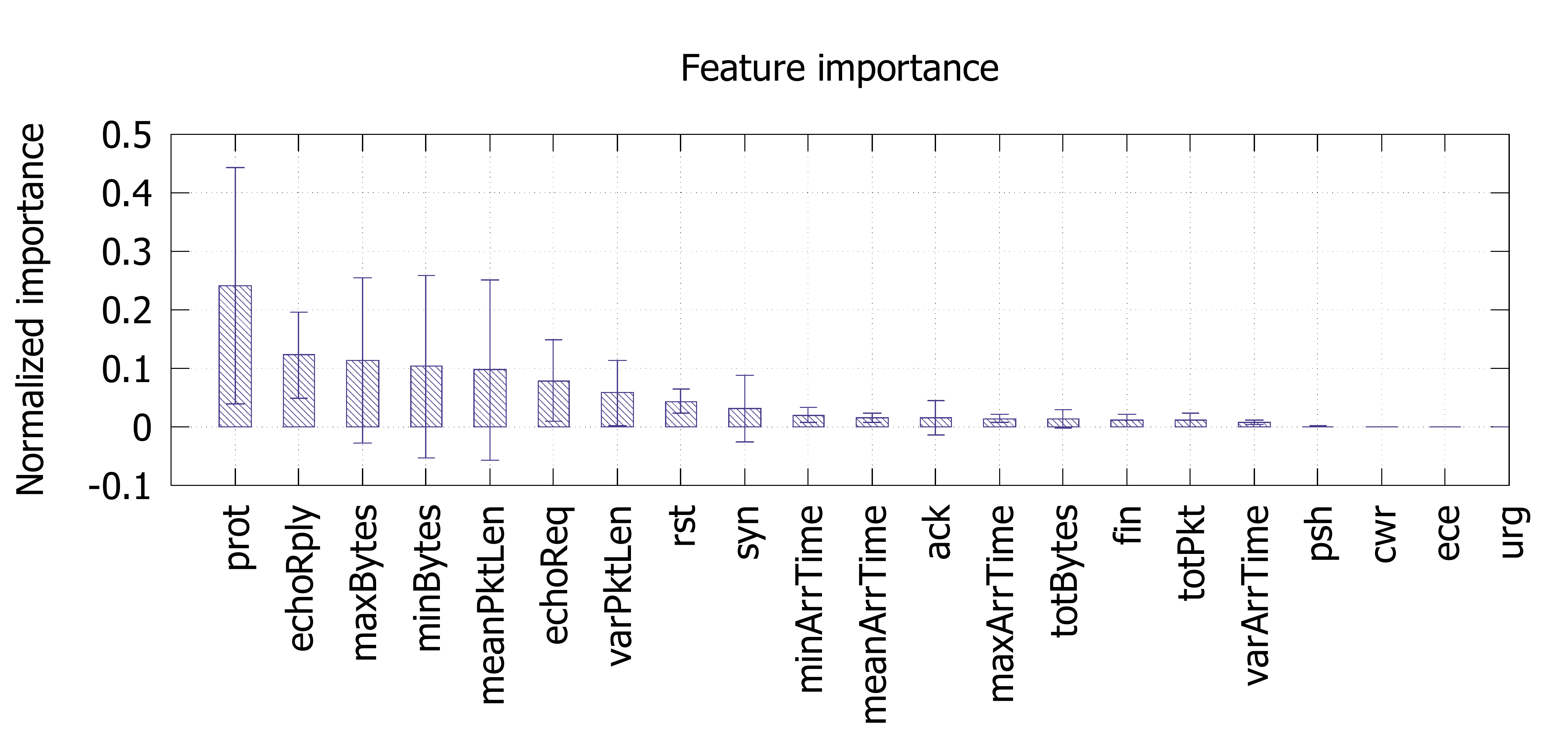}
\caption{Feature ranking in terms of importance}
\label{fig:rank}
\end{figure}

\paragraph{Feature Selection} We gather many features of the network traffic. Few of the features are selected based on their importance ranking to classify the traffic. In Fig. \ref{fig:rank}, features are plotted according to their importance/ranking that can be found using the \emph{feature\_importance} package in the \emph{RandomForest} library. 

\begin{figure}[h]
\centering
\includegraphics[width=\linewidth]{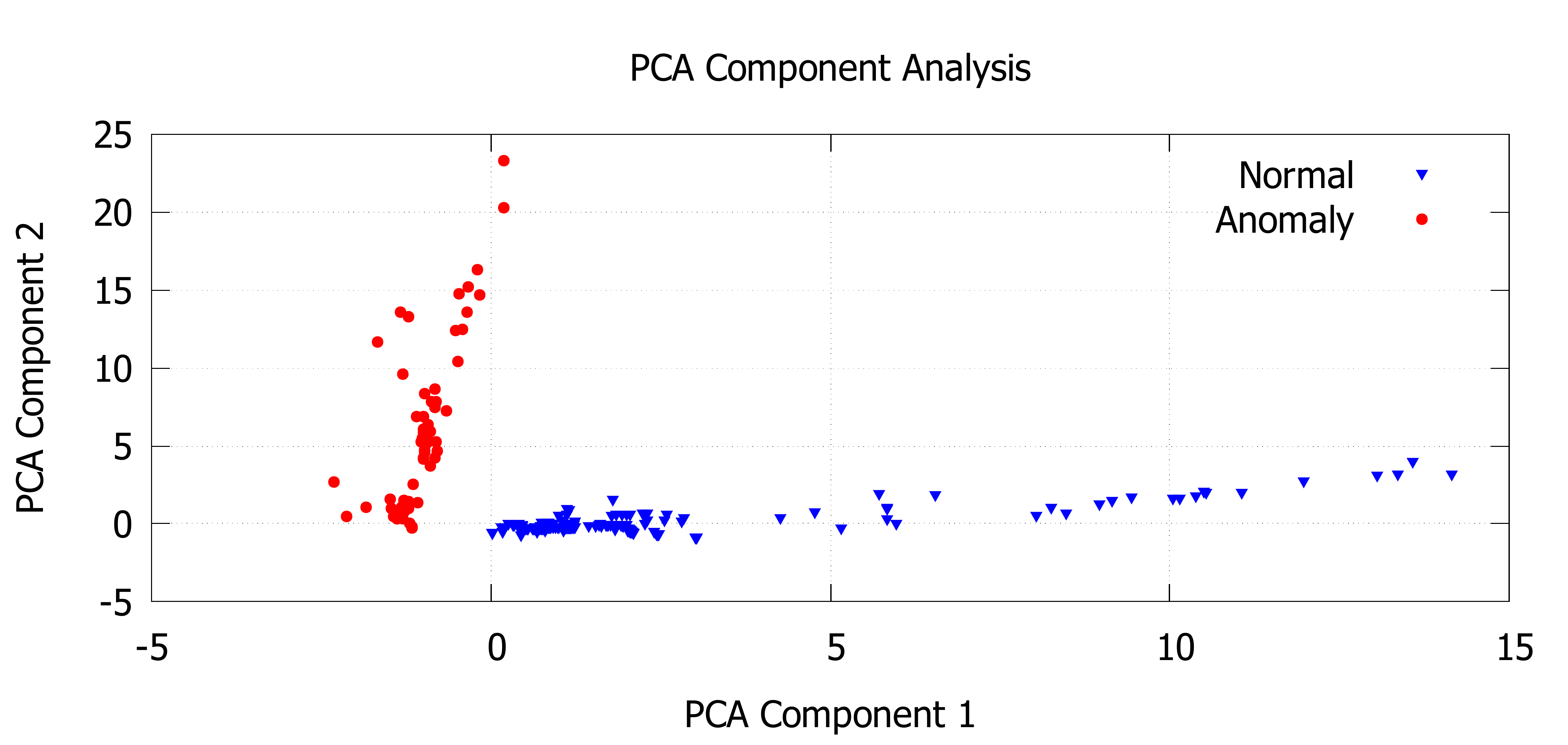}
\caption{PCA component analysis}
\label{fig:AD}
\end{figure}

Additionally, we analyze the features  using Principal Component Analysis (PCA) with two components, that reduces the feature set to a two dimensional space. It is evident from Fig.~\ref{fig:AD} that the data points from \emph{Normal} and \emph{Attacked} traffic are well separated and thus distinguishable.

\paragraph{Traffic Classification} We perform experiments to choose the appropriate classifier in terms of efficiency of attack detection. In this experiment, we consider \emph{SYN attack, ICMP attack} to be distinguished from \emph{normal traffic}. In Table~\ref{tab:conf}, confusion matrix for different classifiers are presented. From the table, we observe that Random Forest is able to classify 626 out of 633 data points as normal traffic. One can see the diagonal elements of the confusion matrix to visualize the efficiency of the classifier. We calculate the accuracy for each of the classifiers as
\begin{equation}
\mathnormal{Accuracy} = \frac{TP+TN}{TP+TN+FP+FN}
\end{equation}
where, \emph{TP}, \emph{TN} and \emph{FP}, \emph{FN} refer to True Positive, True Negative, False Positive and False Negative, respectively.

It is observed that \emph{Random Forest} classifier detects the attacks more accurately among others. This is evident as prior research shows Random Forest gives more accurate results than SVM or Logistic Regression with a multi-class data set~\cite{statnikov2007random}.

\begin{table}[h]
\centering
\caption{Confusion Matrix}
\label{tab:conf}
\begin{tabular}{@{}|l|l|l|l|@{}}
\toprule
                  & Normal    & SYN attack    & ICMP attack   \\ \midrule
\multicolumn{4}{|c|}{\textbf{Linear SVM (Accuracy : 94.444\%)}}    \\ \midrule
Normal            & 622       & 11              & 0               \\ \midrule
SYN attack      & 58        & 201             & 0               \\ \midrule
ICMP attack     & 0         & 0               & 350             \\ \midrule
\multicolumn{4}{|c|}{\textbf{Logistic Regression (Accuracy : 85.587\%)}}    \\ \midrule
Normal            & 631       & 2              & 0               \\ \midrule
SYN attack      & 177       & 82            & 0               \\ \midrule
ICMP attack     & 0         & 0               & 350             \\ \midrule
\multicolumn{4}{|c|}{\textbf{KNN (Accuracy : 95.088\%)}}           \\ \midrule
Normal            & 626       & 7               & 0               \\ \midrule
SYN attack      & 52        & 207             & 0               \\ \midrule
ICMP attack     & 1         & 1               & 348             \\ \midrule
\multicolumn{4}{|c|}{\textbf{Random Forest (Accuracy : 95.4911\%)}} \\ \midrule
Normal            & 626       & 7               & 0               \\ \midrule
SYN attack      & 49        & 210             & 0               \\ \midrule
ICMP attack     & 0         & 0               & 350             \\ \bottomrule
\end{tabular}
\end{table}

\paragraph{Response Time} The response time of the VNF is defined as the time difference between initiation and detection of the attack. We carried out a series of tests and found that the average response time of the VNF is less than a second.

\begin{figure}[h]
\centering
\includegraphics[width=\linewidth]{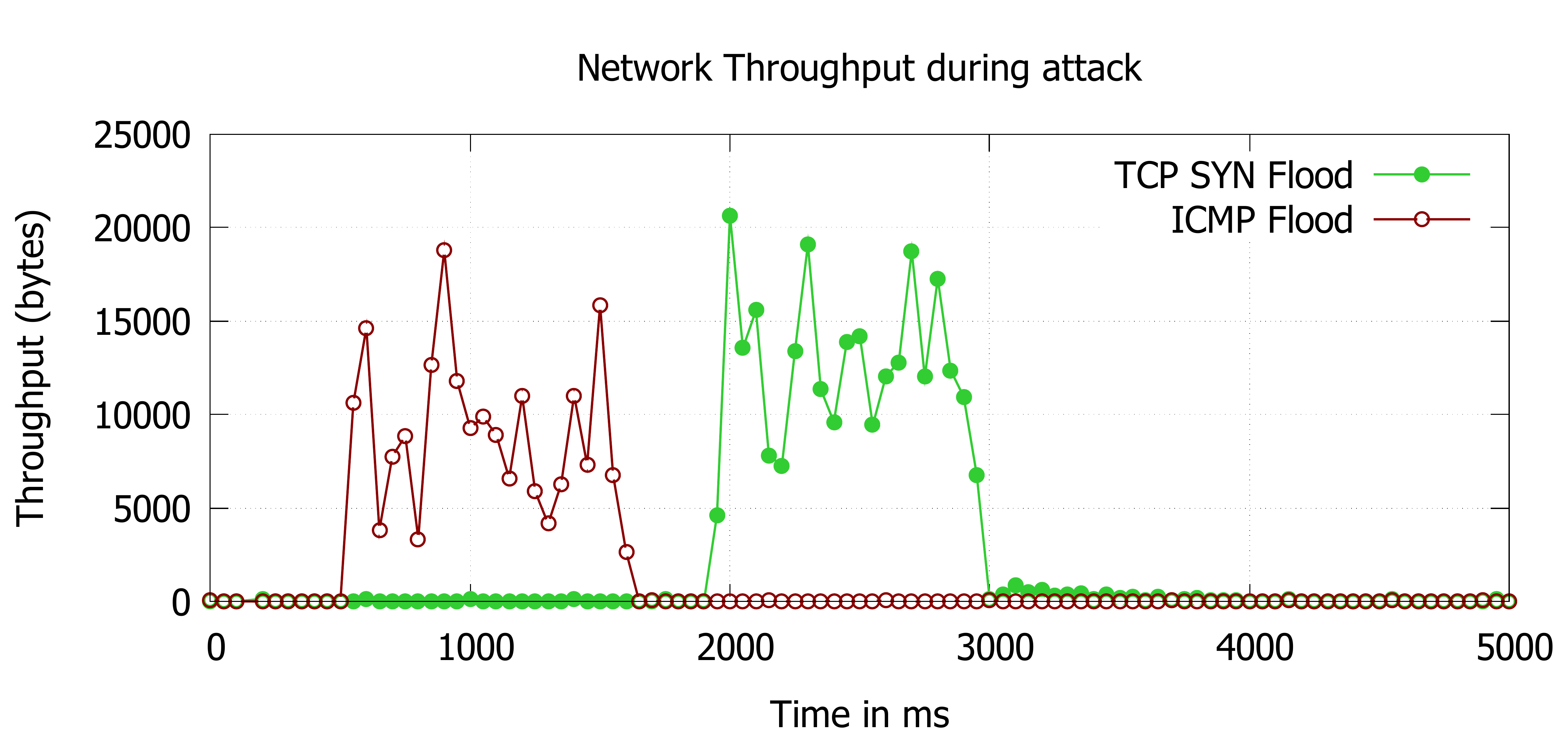}
\caption{Increase in network throughput due to attacks}
\label{fig:throughput}
\end{figure}

\paragraph{Throughput}
During the attack the throughput abruptly increases before detection of the attack and subsequent mitigation. Abrupt increase of throughput (in the scale of thousands of bytes) due to attacks (\texttt{TCP, ICMP} flooding) can be observed in Fig.~\ref{fig:throughput}. Due to this surge in throughput, these attacks can be identified easily and mitigated. The attacks are carried out on IoT devices like TP-Link camera, D-Link smart camera etc. However, after attack mitigation is initiated by concerned VNFs, normal traffic is restored i.e. throughput drops down to hundreds of bytes.  

\vspace{-0.1in}

\section{Related Work}
\label{sec:related}
In this section, we present the existing works in the field of NFV and docker containers. 

In \cite{anderson2016performance}, the advantages of using containers over the existing virtual machine solutions are discussed for virtualizing network components. It is stated that container-based environment provides lower latency and delay variations due to the usage of lower-performing networking schemes. 
Deployment of a container-based virtualized LTE core network is discussed in \cite{fontenla2016lightweight}. The scope of the work was to deploy OpenEPC, an EPC (LTE Evolved Packet Core) implementation using Docker containers.
A framework for deploying VNFs in a light-weight Docker-based container environment is discussed in~\cite{zhang2016opennetvm, zhang2016opennetvm2}. The proposed OpenNetVM is a highly efficient packet processing framework is highly suited for high performance network environments that use complex service chains.
In \cite{cziva2017container}, the opportunities of virtualizing at the network edge is discussed and Glasgow Network Function (GNF), a container-based NFV platform that spawns lightweight container VNFs, is presented. It is shown that this approach saves up the core network utilization and provides a lower latency.
Requirements of IoT edge computing is discussed in~\cite{morabito2018consolidate}. Lightweight virtualization technologies, containers and unikernels, are compared as platforms for enabling scalability, security and manageability in IoT edge computing. In~\cite{pan2018cybersecurity}, various cybersecurity challenges and opportunities of IoT edge computing  are discussed.
An approach for enforcing global security policies in the federated cloud and IoT networks, by implementing the policies on network slices, is described in~\cite{massonet2017end}. NFV and Service Function Chaining (SFC) are used here for implementing the security policies.  
While these works outline the possibility of using lightweight containers, they do not carry out implementation in the context of IoT edge security.
Different from these works, we propose a novel lightweight Docker-based architecture for virtualizing network functions to provide IoT security, implement and carry out performance study.
\vspace{-0.1in}

\section{Conclusion}
\label{sec:conclusion}
In this paper, we proposed \framework - a light-weight Docker-based framework for deploying VNFs at the network edge in order to make NFV compliant with IoT environment. We described how this framework can be applied to enhance the security of IoT. We presented traffic analysis at the network edge for IoT security. This work suggests that NFV will greatly benefit from container-based virtualization. Experimental results have shown that Docker-based VNFs perform well for IoT than existing VM-based frameworks. 
Using the architecture, VNFs that can improve the security of IoT environment were implemented and tested using IoT devices like smart cameras, smart sockets etc. Real time traffic from a TP-Link camera were captured to train the edge-analysis VNF which is able to successfully detect attacks with approximately 95\% accuracy within a second. The known attacks are mitigated using appropriate VNFs, and we will study the handling of \emph{zero-day attacks} in future work. With this work, it is now possible to envisage a scenario where all security VNFs can be deployed at the IoT gateway itself effectively. Our research motivates further investigation in improving security of IoT devices at the network edge with the use of lightweight containers, thus resulting in a smart and secure world of \emph{things}. 
\vspace{-0.1in}


%


\ifCLASSOPTIONcaptionsoff
  \newpage
\fi


\bibliographystyle{IEEEtran}
\bibliography{DockerNFV}
\balance

\end{document}